\documentclass[%
 reprint,
 amsmath,amssymb,
 aps,soul,
pre,
]{revtex4-2}

\usepackage{graphicx}
\usepackage{dcolumn}
\usepackage{bm}
\usepackage{xcolor}

\renewcommand{\selectlanguage}[1]{}
\usepackage{lipsum}
\makeatletter
\newcommand*{\balancecolsandclearpage}{%
  \close@column@grid
  \cleardoublepage
  \twocolumngrid
}
\begin{document}

\preprint{APS/123-QED}

\title{Magnetic reconnection as an Adler-Ohmic bifurcation: \\ The topological origin of Bohm resistivity}


\author{Magnus F Ivarsen}
\email{Contact: magnus.fagernes@gmail.com}
\altaffiliation[Also at ]{
The European Space Agency Centre for Earth Observation, Frascati, Italy}
\affiliation{Department of Physics and Engineering Physics, University of Saskatchewan, Saskatoon, Canada}%


\begin{abstract}
The physical origin of `anomalous’ resistivity in magnetic reconnection remains one of the longest-standing problems in space plasma physics. While the empirical Bohm diffusion scaling ($\eta \propto T/B$) is widely invoked to explain fast reconnection rates, it lacks a rigorous derivation from first principles. Here, we derive this scaling by modeling the ensemble of electron gyro-axes in a magnetized plasma as an overdamped spintronic condensate governed by the Landau-Lifshitz-Gilbert equation. We demonstrate that the breakdown of the ``frozen-in'' condition is rigorously identified as an Adler-Ohmic bifurcation: a topological phase transition where electron gyro-axes lose synchronization with the mean magnetic field. Unlike stochastic turbulence models, this framework predicts a coherent, explosive onset of resistivity that naturally saturates at the Bohm limit. We support this thesis with renormalization group theory and a novel analysis of Magnetospheric Multiscale mission data, which reveals an explosive phase space confinement consistent with collective phase slippage rather than chaotic scattering. These results suggest that Bohm resistivity is a universal topological property of magnetized matter at the critical point of reconnection.
\end{abstract}


\maketitle

\section{Introduction} \label{sec:intro}

Magnetic reconnection is the reordering of magnetic topology, which converts large amounts of magnetic energy into kinetic energy \cite{cowleyTUTORIALMagnetosphereIonosphereInteractions2000,kulsrudMagneticReconnectionSweetParker2001,milanMagneticFluxTransport2007,egedal_large-scale_2012}, thereby powering many of the universe's most energetic transients. Reconnection likewise governs the interaction between Earth's magnetosphere and the solar wind \cite{fujiiControlIonosphericConductivities1987,brunoSolarWindTurbulence2013}. The critical reconnection physics occurs within the electron diffusion region, where the plasma decouples from the magnetic field lines, allowing the lines to snap and reconfigure, and the trigger mechanism for this critical state remains mathematically elusive in magnetohydrodynamic theory.

Our core premise is that magnetic reconnection can be understood as a thermodynamic phase transition. The merits of this position were recently proven by Ref.~\cite{jara-almonte_thermodynamic_2021}, where the authors demonstrated that, for collisional electrons modelled with kinetic (particle-in-cell) simulations, the reconnection trigger is essentially a temperature threshold, and that the system's entropy increased dramatically at the transition. In fact, for collisionless reconnection to occur, the electron distribution function must become agyrotropic \cite{hesse_diffusion_1999,egedal_large-scale_2012}, observations of which have been directly linked to the breakdown of the guiding center approximation \cite{scudder_fingerprints_2002,scudder_frozen_2015}, caused by the width of the thermal distribution.

In general, the nature of the reconnection trigger is debated, and 'anomalous' resistivity, caused by turbulent or stochastic conductivity, is deemed necessary \cite{tsunetaStructureDynamicsMagnetic1996}. Such anomalous resistivity has, throughout the space age, been required to account for the fast dissipation rates observed in magnetic reconnection events, empirically described using the Bohm diffusion scaling ($\sim T/B$) \cite{kaufmanExplanationBohmDiffusion1990}.

In this article, we derive Bohm resistivity from first principles by treating the magnetized electron fluid as an overdamped spintronic condensate governed by the Landau-Lifshitz-Gilbert  equation \cite{lakshmanan_fascinating_2011}, representing the breakdown of ideal magnetohydrodynamics. Following renormalization group theory (which we apply in Appendix~B), we find that at the critical point, the physics of reconnection reduces to the topological constraints of the stable electron gyro-oscillations. To bolster the position, we demonstrate that magnetic reconnection acts as an Adler-Ohmic bifurcation, and the Ohmic response of the system naturally recovers Bohm resistivity. The implication is that the 'frozen-in' condition (or guiding center approximation) obeys a tilted washboard potential, mathematically identical to phase locking in Josephson junction theory.

While magnetic topology is traditionally viewed as a non-local constraint, Ref.~\cite{mactaggart_field_2025} has recently demonstrated that reconnection rates can be rigorously derived from local field line slippage metrics, validating the notion that the reconnection trigger can be modeled as a local synchronization failure (phase slip). In so doing, we shall defend the position that the 'anomalous' resistivity in magnetic reconnection is the  macroscopic signature of the electron gyro oscillations failing to maintain coherence against sharp magnetic gradients, thereby spiraling into chaotic orbits.


By explicitly modeling the topological skeleton constituted by the electron gyro-axis sensemble, we are able to describe the natural damping of the gyro-axes against the mean magnetic field, and we find that the breakdown of this synchrony, the failure of the ``frozen-in'' condition, is governed by the universal symmetries of the order parameter. Consequently, we model the system as a `spintronic condensate,' a proxy for the XY universality class, and we thereby argue that the macroscopic laws of reconnection are agnostic to the model's microscopic realization. The breakdown of magnetohydrodynamics is thereby understood as a Berezinskii-Kosterlitz-Thouless (BKT) transition.
 
In Section~II we summarize the theoretical foundation behind the gyro-axis slippage model (whose details are contained in Appendix~A), demonstrating that Bohm resistivity follows from the model's premises. Then, in Section~III we present first simulations of the field-perpendicular plane, and then a novel analysis of phase space bunching in observations by the Magnetic Multiscale (MMS) mission. In section~IV, bolstered by Appendix~B we argue that magnetic reconnection can be considered a critical phase transition where electron gyro-axes collectively slip away from the forces that keep them aligned with the magnetic field. In section~V we apply renormalization group (RG) theory to the equation of motion with inertia, documenting that the macroscopic-scaling of the model is overdamped. In section~VI we demarcate area of validity and offer an outlook.


\section{Methodology}

Our goal is to define a minimal model of electron gyro oscillations. For the purposes, we model the electron's axis of gyration as a rotor that aligns with the local mean field, for which we must define the geometry of alignment on the unit sphere.

We define a lattice of overdamped rotors subject to three competing torques: intrinsic chirality, ferromagnetic exchange, and thermal agitation. The resulting Langevin equation describes the time-evolution of the orientation vector $\hat{n}_i$ \cite{ivarsen_information_2025},
\begin{equation} \label{eq:eom3d}
\frac{\partial\hat{n}_{i}}{\partial t}= (\hat{\omega}_{i}\times\hat{n}_{i}) + \zeta[\hat{n}_{i}\times(\hat{\Psi}_{i}\times\hat{n}_{i})] + \eta_{i}(t).
\end{equation}
Here, $\hat{\omega}_{i}$ represents the intrinsic drive (spin precession), $\Psi_{i}$ captures the local alignment field from neighbors, and $\eta_{i}$ introduces stochastic (Gaussian) noise fluctuations.

\begin{figure*}
    \centering
    \includegraphics[width=\textwidth]{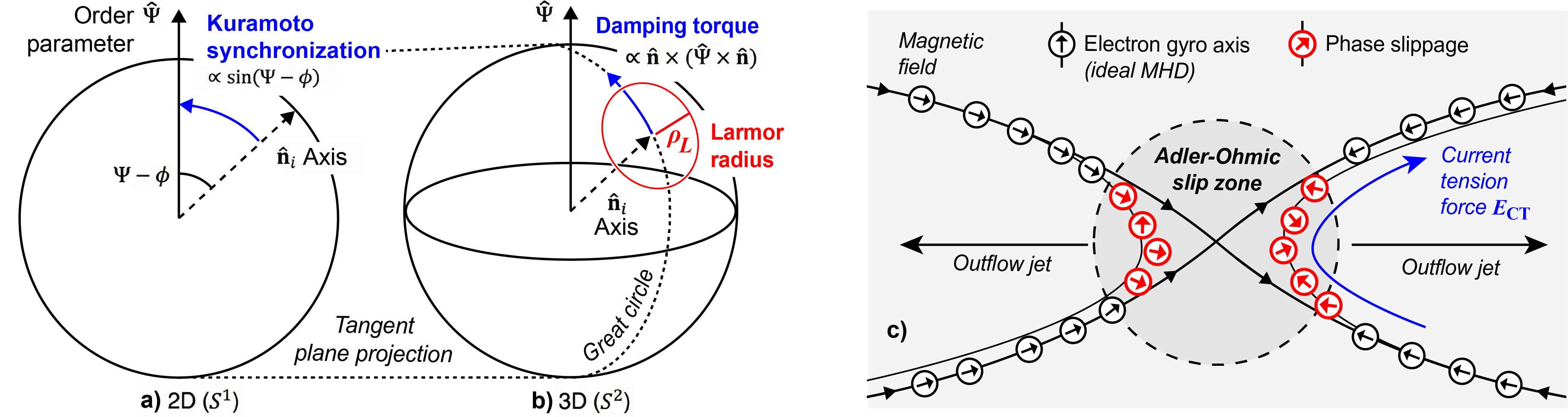}
    \caption{\textbf{Panel a):} the geometry of the two-dimensional tangent plane of the electron gyro rotor, showing how the restoring force acts to align the rotor's axis with the mean field $\Psi$. \textbf{Panel b):} the geometry of the full three-dimensional model, where the electron gyro-axis is the unit vector on the unit sphere ($|\hat{n}_i|=1$), and where the vector torque proportional to $(\hat{n}_i \times \hat{\mathbf{\Psi}}) \times \hat{n}_i$ (Eq.~\ref{eq:eom3d}) acts to align the rotor with the mean field  \cite{lakshmanan_fascinating_2011}. \textbf{Panel c):} Schematic representation of a reconnection event in the overdamped spintronic condensate (black and red compasses representing rotors that align with the magnetic field). The current tension electric field is indicated, which causes slipping electrons to enter the outflow jets.}
    \label{fig:geometry}
\end{figure*}

The force interaction is defined by the requirement for rotor $i$ to align with its local neighborhood $\mathcal{N}_i$, analogous to the Heisenberg ferromagnet \cite{vicsek_novel_1995,toner_flocks_1998,marchetti_hydrodynamics_2013}. The local field $\boldsymbol{\Psi}_i$ is the vector sum of neighbors:
\begin{equation}
    \boldsymbol{\Psi}_i = \sum_{j \in \mathcal{N}_i} \hat{n}_j.
\end{equation}
Because the rotors are constrained to the unit sphere ($|\hat{n}|=1$), the alignment torque is the projection of this field onto the rotor's tangent plane, recovering the Landau-Lifshitz torque \cite{lakshmanan_fascinating_2011} (see also Ref.~\cite{lohe_non-abelian_2009}),
\begin{equation}
    \boldsymbol{\mathcal{T}}_\text{LL} = \zeta \;\hat{n}_i\times (\boldsymbol{\Psi}_i \times \hat{n}_i).
\end{equation}
As illustrated in Figure~\ref{fig:geometry}, the magnitude of this torque projects to the two-dimensional alignment plane as a Kuramoto-coupling,
\begin{equation}
    |\boldsymbol{\mathcal{T}}_\text{LL}| = \zeta \sin(\Psi_i-\phi_i),
\end{equation}
establishing that the relevant degree of freedom is the phase mismatch $\delta\phi\equiv\phi_i-\boldsymbol{\Psi}_i$, allowing us to separate the dynamics into an effortless precession around $\boldsymbol{\Psi}_i$ (Goldstone mode) and a dissipative relaxation of the phase mismatch.

The torque $\boldsymbol{\mathcal{T}}_\text{LL}$ drives the time derivative of the phase mismatch, yielding \cite{strogatz_coupled_1992}:
\begin{equation} \label{eq:adler}
    \dot{\delta\phi_i} = \omega_i - \zeta \sin(\delta\phi_i),
\end{equation}
where $\omega_i$ is the scalar magnitude of the intrinsic chirality (mismatch). Eq.~(\ref{eq:adler}) is recognized as the Adler equation for phase-locking oscillators (Josephson junctions) \cite{danner_injection_2021}. This allows us to define the gyro-axis slip velocity,
\begin{equation} \label{eq:vslip2_m} 
    v_\text{slip} \equiv \langle \dot{\delta\phi}\rangle = \sqrt{\omega_i^2-\zeta^2},
\end{equation}
where a total slippage of $\pi$ yields field-reversal.

By performing a Taylor expansion of the model's local field around the rotor (see Appendix~A), we recover the Landau-Lifshitz-Gilbert equation \cite{lakshmanan_fascinating_2011},
\begin{equation}\label{eq:llb_m}
   \frac{\partial \hat{n}_i}{\partial t} = -\omega_{i} \xi^2 (\hat{n}_i \times \mathbf{H}_{\text{eff}}) - \zeta \xi^2 \hat{n}_i \times (\hat{n}_i \times \mathbf{H}_{\text{eff}}),
\end{equation}
where chirality is provided by the electron cyclotron frequency $\Omega_{ce}$, via $\omega_i\to\Omega_{ce}+\delta\omega$, and where the correlation length equals the electron inertial length, $\xi \approx d_e$. By recognizing that $\langle\omega_i\rangle=\Omega_{ce}$, we observe that a broadening of $\omega_i$, which drives the model's dynamics \cite{ivarsen_onsager_2025-1,ivarsen_information_2025,ivarsen_kinetic_2025}. This is the thermal speed of the electrons $v_{\text{th},i}$, which yields $\omega_i = v_{\text{th},i}/L_{\nabla B}$, $L_{\nabla B}$ being the length-scale of the magnetic gradient \cite{goldston_introduction_1995}. This allows us to define the plasma Adler equation,
\begin{equation}
\label{eq:plasmaadler_m} 
\dot{\delta\phi}_i = v_{\text{th},i} \left( \frac{1}{L_{\nabla B}} - \frac{1}{\rho_L}\sin(\delta\phi_i) \right),
\end{equation}
where $\Omega_{ce}=v_{\text{th},i}/\rho_L$, $\rho_L$ being the Larmor radius of the precessing electron  \cite{stasiewicz_finite_1993}. Eq.~(\ref{eq:plasmaadler_m}) defines the tilted washboard potential in terms of adiabatic invariance and the breakdown of stable electron orbits \cite{buchner_regular_1989}. This recovers the magnetohydrodynamic threshold for magnetic reconnection,
\begin{equation} \label{eq:trigger0_m}
    \rho_L > L_{\nabla B}.
\end{equation}

This threshold is conventionally derived through considering the magnetic moment $\boldsymbol{\mu}$ as no longer conserved (the breakdown of adiabatic invariance), and its recovery from the model, essentially a consideration of vector alignment on the unit sphere, provides justification for the model's application to magnetized plasmas undergoing magnetic reconnection. 

During gyrophase slippage (a rapid loss of adiabatic electron orbits), electrons collectively slip against the magnetic topology, generating an effective electric field that feeds the reconnection jets (see Figure~\ref{fig:geometry}c). This slippage electric field (blue arrow) is physically equivalent to the inertial field created by current tension \cite{luo_current_2024}. When the current density exceeds the critical threshold ($J>J_c$), the topological constraint breaks, triggering a collective phase slip that accelerates particles into the outflow.


As we detail in Appendix~A, by quantifying the velocity of the phase-slipping gyro-axes, we can describe the work performed by the effective electric field associated with the slippage drift. Amp\`{e}re's law eventually yields the resistivity,
\begin{equation} \label{eq:etaspin_m}
\eta_{spin}(J) = \eta_0 \sqrt{1 - \left(\frac{J_c}{J}\right)^2} \quad \text{for } J > J_c,
\end{equation}
where the characteristic resistivity scale is $\eta_0~=~\mu_0 \rho_L v_{\text{th},i}$, and where,
\begin{equation}
    J_c = \frac{e B^2}{m \mu_0 v_{\text{th},i}} = \frac{B}{\mu_0\rho_L},
\end{equation}
is the critical current, with current $J = |\nabla\times \mathbf{B}|/\mu_0$ being caused by the sharp field gradients. The characteristic resistivity scale simplifies by noting that $\rho_L v_{\text{th},i} \approx k_B T/e B$, meaning that $\eta_0 \propto B^{-1}$. This recovers the Bohm resistivity scaling \cite{braginskiiTransportProcessesPlasma1965,ottDiffusionStronglyCoupled2011}), a standard phenomenological relation invoked in space \& laboratory plasmas \cite{kaufmanExplanationBohmDiffusion1990,treumann_advanced_1997,burch_electron-scale_2016}.

\section{Results}

\subsection{Simulations}

To provide empirical evidence for the model's theoretical description, we simulate the system numerically, by assigning discrete lattice sites to renormalized coherence volumes scaled to the electron inertial length ($\xi \approx d_e$). At the critical point, the correlation length dominates, allowing us to treat these fluid elements as discrete, rigid rotors interacting via a mean field, to which they dampen with a Landau-Lifshitz torque. This discretization, which we describe below, is the physical manifestation of the finite Larmor radius threshold.

We simulate the two-dimensional tangent plane dynamics, obtained by solving the Adler equation (Eq.~\ref{eq:adler}). This $S^1$ unit circle corresponds to the two-dimensional cross-section of the plasma, the current sheet perpendicular to the  magnetic field $\mathbf{B}$. As alluded to, the spintronic condensate is treated as a discrete lattice of $N \times N$ coupled electron gyro-axes, with $N=512$ ($2.6\times10^5$ rotors), where the state variable is the instantaneous phase of the gyro-axis (the rotor) $\phi_{ij} \in [0, 2\pi)$, in the tangent plane projection of the alignment torque (Figure~\ref{fig:geometry}), with $\Psi-\phi$ being the polar phase mismatch between the rotor and the local mean field.

From Eq.~(\ref{eq:adler}) we observe that the dynamics of the evolving phases are governed by the overdamped Adler equation in the co-rotating Larmor frame (setting $\langle \Omega_{ce} \rangle \equiv 0$), yielding a Kuramoto model \cite{ivarsen_information_2025},
\begin{equation}\label{eq:kuramoto}
    \frac{d\phi_i}{dt} = \delta\omega_i(T_e) + K R_i \sin(\Psi_i - \phi_i) + \eta_i(t),
\end{equation}
where $R_i$ and $\Psi_i$ are the amplitude and phase of the local order parameter, defined by the spatial average over the neighborhood $\mathcal{N}_i$,
\begin{equation}
    R_i e^{i\Psi_i} = \langle e^{i\phi_j} \rangle_{j \in \mathcal{N}_i},
\end{equation}
In our simulations, this average is computed using a Gaussian kernel with width $\sigma = 2.0$ (in units of lattice spacing), representing the exchange interaction length scale $\xi \approx d_e$.

\begin{figure*}
    \centering
    \includegraphics[width=\textwidth]{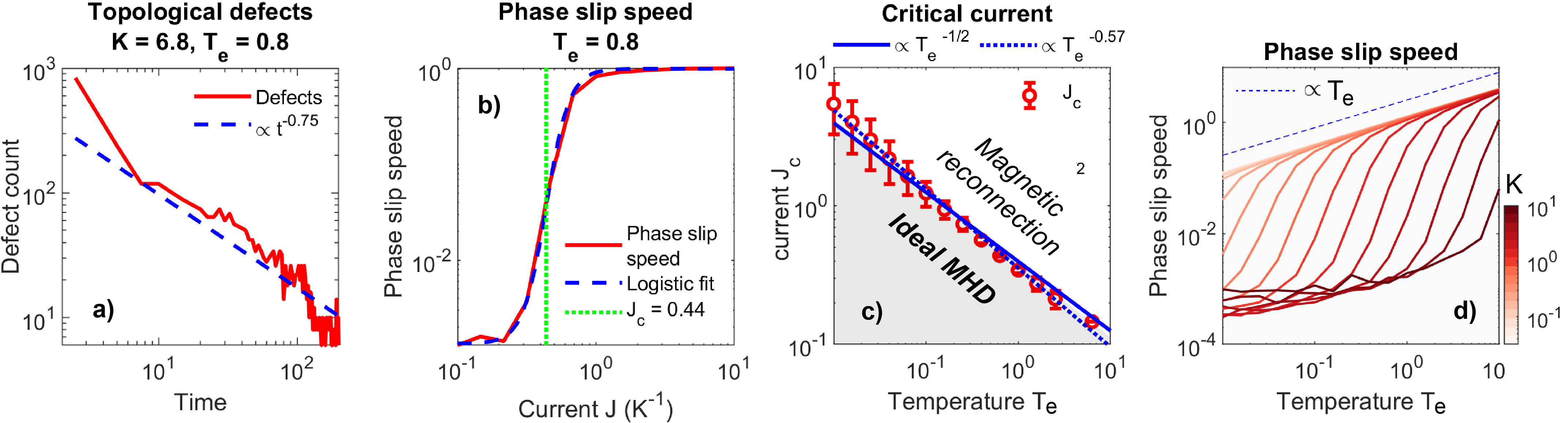}
    \caption{A statistical aggregate of 512 simulation runs, each featuring a two-dimensional $512\times512$ lattice, systematically varying $0.1<T_e<3.2$ and $0.032<K<10$ in logarithmic increments.\textbf{ Panel a)} shows the defect count (red) and a $t^{-0.75}$ powerlaw fit (blue dashes). \textbf{Panel b)} shows the ensemble average phase slip speed (red, evaluated from the second half of 16 simulations with varying values of $K$), a logistic fit (blue dashes), and $J_c$, the deflection point in the logistic curve (green dots).  \textbf{Panel c)} shows $J_c$ for all the simulations, in bins of temperature $T_e$. Errorbars denote the 95$^\text{th}$ percent confidence interval of the logistic fit in each bin. Fits of $\propto T_e^{-0.57}$ (determined through non-linear least squares minimziation) and $\propto T_e^{-1/2}$ are shown in blue dashes and blue solid line respectively; the shaded gray region delineates  ideal magnetohydrodynamics from magnetic reconnection. \textbf{Panel d)} shows ensemble average phase slip speed, varying the temperature (the color of the line indicates the coupling strength $K$). A $\propto T_e$ fit is shown in blue dashes. }
    \label{fig:stats}
\end{figure*}

In Eq.~(\ref{eq:kuramoto}), $K$ is the coupling constant, and $\delta\omega$ is the temperature-driven ``perturbation'' to the electron cyclotron frequency, evaluated as a Gaussian random variable with distribution width equal to $T_e$. We performed $16\times16=512$ simulations, systematically varying $0.1<T_e<3.2$ and $0.032<K<10$ in logarithmic increments. This entails varying the ratio $\Gamma=K/T_e$, coupling stiffness divided by noise: $10^{-2}<\Gamma<10^2$, where transitions are expected around $\Gamma\approx1$. 

From each simulation run, we output two key quantities, \textit{(1)} topological defect count $N_d$, equal to the number of nonzero topological winding numbers, $q$,
\begin{equation}
    q = \frac{1}{2\pi} \oint_C \nabla \phi \cdot d\mathbf{l},
\end{equation}
evaluating the entire lattice, and \textit{(2)} the ensemble average phase slip speed $v_\text{slip} = \langle|\dot{\phi}-\dot{\Psi}|\rangle$, the rate at which the electrons' magnetic moments de-synchronize from the magnetic field. An aggregate of these quantities are shown in Figure~\ref{fig:stats}.

Phase synchronization leads to vortices, which merge by synchronizing neighbour rotors. However, Figure~\ref{fig:stats}a) shows that the decay of the defect number density $N_d(t)$ follows a $t^{-0.75}$ scaling \cite{carnevale_evolution_1991,larichev_weakly_1991}, markedly slower than the diffusive $t^{-1}$, and therefore indicative of anomalous transport, consistent with magnetic island coalescence instability in magnetohydrodynamics \cite{dorelli_electron_2001}. Turbulence in our model thereby relaxes via a hierarchical merger of magnetic flux tubes. 

Figure~\ref{fig:stats}b) shows the ensemble phase slip speed, for simulations that sweep through coupling $K$, where we plot against current $J=K^{-1}$, wherein the angular momentum required to decouple the gyro-axis from the mean field ($\omega_i+\eta_i(t)$) is opposed by coupling strength $K$ (Eq.~\ref{eq:kuramoto}), and where we use the magnetohydrodynamic consideration of current as driven by magnetic stress $J=\nabla\times\mathbf{B}/\mu_0$. A logistic fit of $v_\text{slip}$ against $J$ reveals that the onset of phase slippage is a continuous phase transition with an explosive increase at $J_c$, the critical current. 

Figure~\ref{fig:stats}c) shows the critical current $J_c$, calculated from 16 logistic plots of 16 ensemble simulations that each sweep through $K$, sorted by $T_e$ (red circles with errorbars). We observe a near-perfect anti-correlation, with a $\propto T_e^{-1/2}$ fit adequately describing the data, implying that $J_cT_e=constant$. This means that the critical coupling strength required to maintain the phase-locked state scales inversely with the temperature, or, equivalently, the topological lattice breaks when the thermal frustration energy ($k_BT_e$) exceeds the Josephson binding energy of the rotors. The straight line in Figure~\ref{fig:stats}c) thereby delineates ideal magnetohydrodynamics from magnetic reconnection.

Figure~\ref{fig:stats}d) shows 16 curves of $v_\text{slip}$ against temperature $T_e$, with a $\propto T_e$ fit shown with blue dashes, showing that when the system enters into phase slippage at the Adler-Ohmic crossover, which happens at hotter temperatures for higher coupling strengths, it does so with an Ohmic scaling.

\begin{figure}
    \centering
    \includegraphics[width=.5\textwidth]{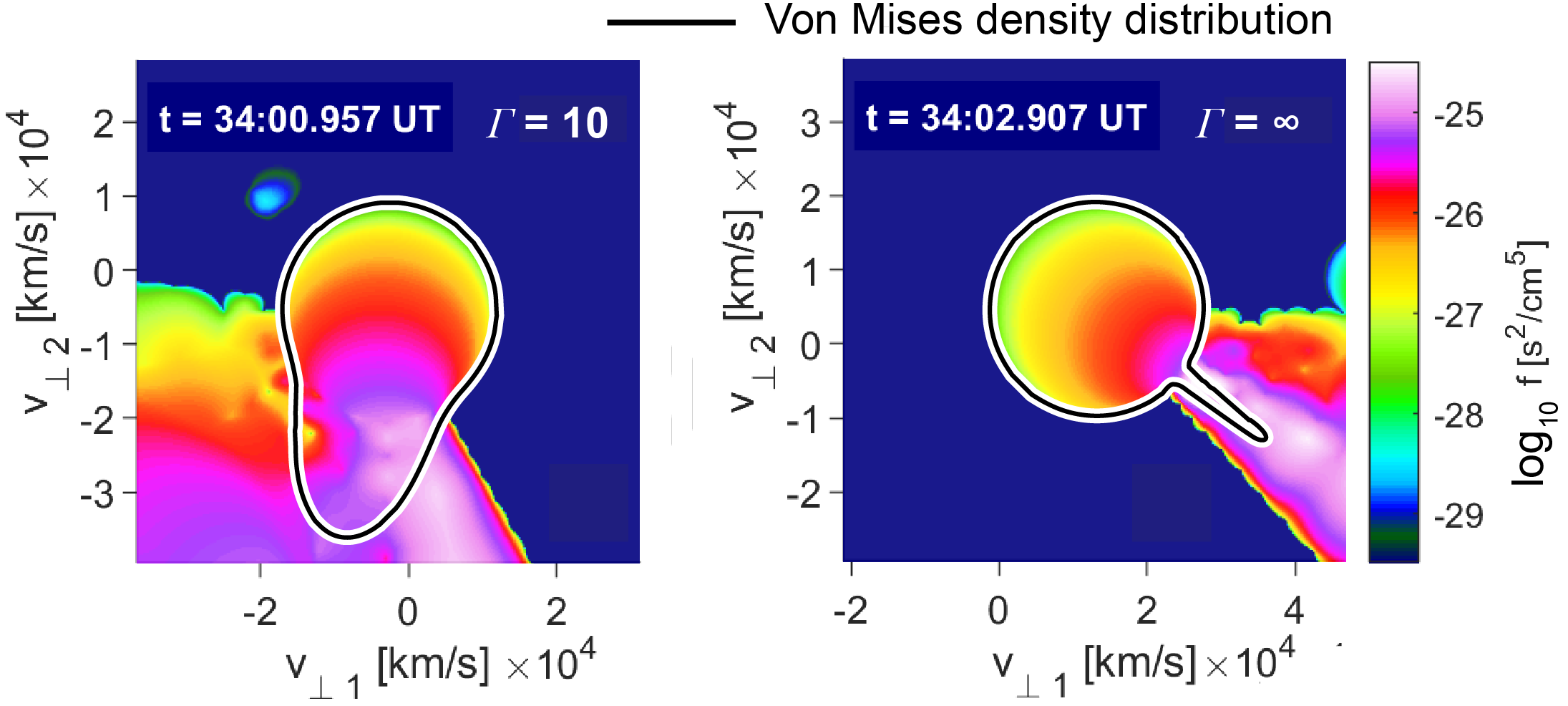}
    \caption{\textbf{Electron agyrotropy observed by MMS-3 during a magnetotail reconnection event.} Both panels shows perpendicular electron velocity phase space cross-sections with a colorscale (taking the $v_\parallel\approx0$ cross-section). Plots of Von Mises density distributions (Eq.~\ref{eq:mises}) fitted to the $5000$~km/s~$<v_\perp<~25,000$~km/s electron population are overlaid on top of a circle encompassing the thermal ($v_\perp<17,000$~km/s) population. The velocity phase portraits are shifted by the measured perpendicular bulk flow.}
    \label{fig:mms0}
\end{figure}

\subsection{Observational evidence}

There is observational evidence for our thesis. Electron velocity phase space `crescents', or gyrophase bunching, have been widely observed in high-resolution (burst-mode) observations of magnetic reconnection by the Magnetospheric Multiscale (MMS) mission \cite{burch_electron-scale_2016,torbert_electron-scale_2018}. Furthermore, agyrotropy is a consistent signature of magnetic reconnection \cite{aunai_electron_2013,bessho_electron_2014,swisdak_quantifying_2016}. Ref.~\cite{cohen_dominance_2017} notably observed that electrons maintained phase coherence over thousands of kilometers, directly tying this synchronization to the underlying magnetic topology.

\begin{figure*}
    \centering
    \includegraphics[width=\textwidth]{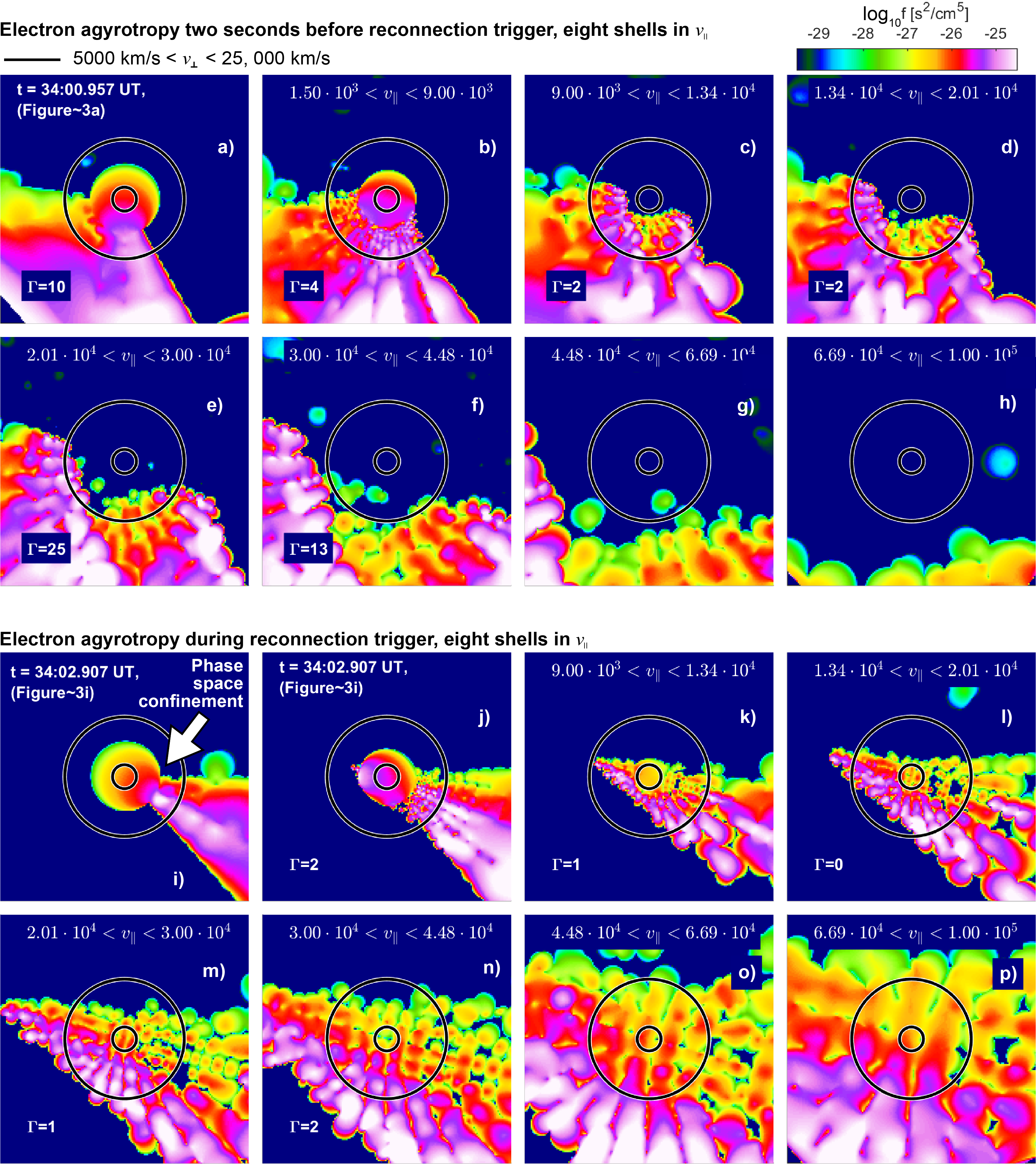}
    \caption{\textbf{Perpendicular suprathermal electron velocity phase portraits in successive shells of $v_\parallel$.} Panels~a--h) show a snapshot  two seconds prior to the reconnection trigger-point ($J\approx100$nA/m$^2$), while as Panels~i--p) show the reconnection trigger-point ($J\approx10$nA/m$^2$). Panels a) and i) correspond to Figure~\ref{fig:mms0}a) and i). The definition of the suprathermal population is indicated between to black circles. The globular pattern shown is likely affected by the instrument's geometry being ``filled'' with electron phase space density $f$.}
    \label{fig:mosaic2}
\end{figure*}

\begin{figure*}
    \centering
    \includegraphics[width=0.86\textwidth]{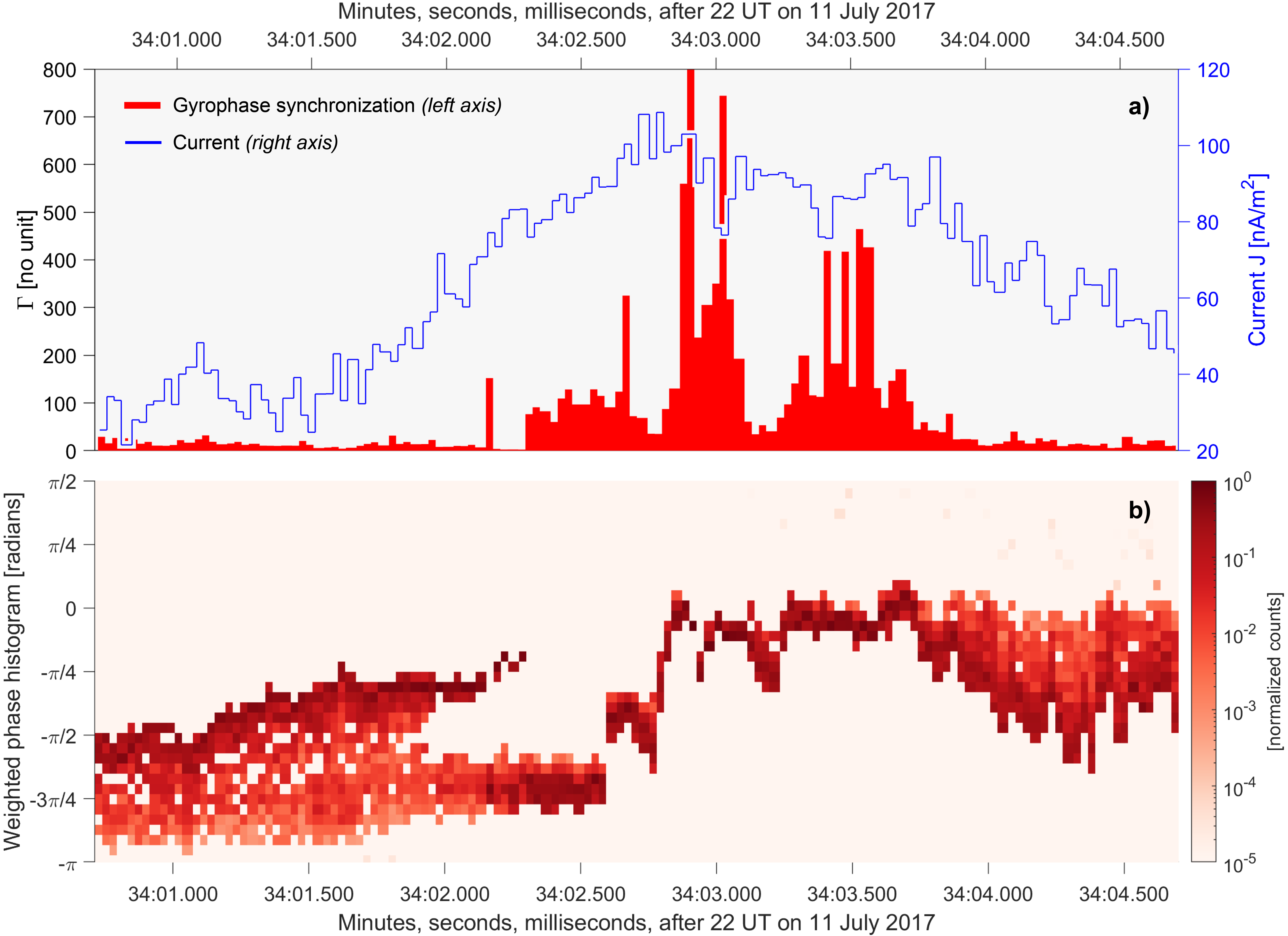}
    \caption{\textbf{MMS-3 observations for the magnetotail reconnection event presented in Ref.~\cite{torbert_electron-scale_2018}.} \textbf{Panel a)} shows the time series of the synchronization order parameter $\Gamma$ (Eq.~\ref{eq:mises}, left axis) and current density (right axis). \textbf{Panel b)} shows the time-resolved weighted phase histogram for the suprathermal electron population ($5,000 < v_{\perp} < 25,000$ km/s). The color scale represents the accumulated phase space density ($\Sigma f$) within each azimuthal bin. We note that the valley in $\Gamma$ around 34:03.250~UT is likely caused by the spacecraft crossing the separatrices, or boundaries, of the reconnection exhaust \cite{torbert_electron-scale_2018}, and that the peak current  $J_{max} \approx 105$ nA/m$^2$ suggests a critical threshold for the bifurcation that is consistent with saturation levels reported in kinetic reconstructions of this event \cite{korovinskiy_inertia-based_2021}.
    }
    \label{fig:mms}
\end{figure*}

To explicitly link such observations to electron gyro-axis slippage, we analyze the magnetic reconnection event presented by Ref.~\cite{torbert_electron-scale_2018}, processing field observations from the MMS-3 probe as it crossed the X-line of a magnetotail reconnection event at 22:34 UT on 11 July 2017. We transform electron velocity distribution functions into the field-aligned fluid rest frame by subtracting the perpendicular bulk velocity, thereby isolating intrinsic phase structure from kinematic drift. We calculate the local current density $J$ directly from the electron fluid moments ($J \approx -e n_e \mathbf{v}_e$), providing a high-resolution proxy for the local magnetic stress driving the event. Finally, we isolate the trapped suprathermal population (the crescent features observed by Ref.~\cite{torbert_electron-scale_2018}), demanding $|v_{\parallel}| < 1500$ km/s, $E > 250$~eV, and we quantify the degree of gyrophase synchronization by fitting the perpendicular angular distribution to the Von Mises density function \cite{fernholz_von_2012},
\begin{equation} \label{eq:mises}
P(\phi|\Gamma) = \frac{e^{\Gamma \cos(\phi - \mu)}}{2\pi I_0(\Gamma)},
\end{equation}
where $I_0$ is the 0$^\text{th}$-order Bessel function, and where the extracted concentration parameter $\Gamma$ serves as the scalar topological order parameter, akin to the pressure tensor off-diagonal terms $\sqrt{Q}$ utilized by Refs.~\cite{aunai_electron_2013,swisdak_quantifying_2016}. Eq.~(\ref{eq:mises}) allows us to quantify the sharp phase-locking features observed by Refs.~\cite{burch_electron-scale_2016,torbert_electron-scale_2018}.

High values of $\Gamma$ indicate that the gyrophases in the electron diffusion region are strictly synchronized (anisotropic). Whereas ideal MHD predicts a disordered gas phase with $\Gamma\approx 0$ (an isotropic Maxwellian distribution \cite{gurnett_introduction_2017}), the observation of $\Gamma\gg 1$ reveals the formation of a topologically locked \emph{solid}.

In Figure~\ref{fig:mms0}, we present the result of the foregoing analysis. The nine panels show time-resolved phase space cross-sections of the perpendicular electron velocity distribution ($v_{\parallel} \approx 0$) leading up to the reconnection event, with the critical trigger-point presented in panel i). In panel i), we observe topological phase space confinement in the form of a sharply defined funnel, with $\Gamma=\infty$, far exceeding the angular resolution of the instrument. 

Figure~\ref{fig:mosaic2} shows perpendicularly suprathermal electrons, in successive shells of parallel velocity $v_\parallel$, segmenting the data in logarithmically spaced bins. The first bin represents the field-perpendicular electrons $v_\parallel\approx0$, and we observe that the successive bins do not exhibit a sharp Von Mises  distribution and therefore do not exhibit collective motion (Bohm diffusion) akin to the purely field-perpendicular electrons. These, which substantial parallel velocities, will exhibit phase-mixing caused by the rapidly changing magnetic field, and will exit the region fast. 

In Figure~\ref{fig:mms}a) we show the complete time series $\Gamma(t)$ (red bars) for the purely perpendicular suprathermal electrons, with the current $J$ overlaid. Prior to the critical transition (before 34:02.800~UT), $\Gamma$ remains low, though non-zero. This baseline stiffness reflects the natural agyrotropy of the trapped suprathermal plasma and residual drift gradients in the inflow. As the current (blue line) ramps up toward a maximum of $J_{max} \approx 105$~nA/m$^2$, $\Gamma$ rises abruptly, reaching saturation levels $>800$ (values approaching singular synchronization $\Gamma \to \infty$ are truncated to 800 for visualization). This explosive rise coincides exactly with the current peak, after which the current saturates and decreases. By 34:04.000~UT, the system relaxes and $\Gamma$ returns to baseline.
The striking synchronization of the detected electrons during the trigger point, which we show in Figure~\ref{fig:mms}b) through weighted histograms of phase space density, demonstrate collective transport across field-lines rather than chaotic scattering behaviour.

\section{Discussion~A}

Our model simply states that the electron gyro-axis, modeled as a rotor in three dimensions, aligns, via a torque, with the mean field. This is the \textit{fundamental microscopic constraint} of ideal magnetohydrodynamics, manifesting the principle of adiabatic invariance \cite{cowling_magnetohydrodynamics_1957}.  We model the slippage of this alignment, essentially the failure of the first adiabatic invariant during magnetic reconnection.

The foregoing $S^2$ dynamic in three dimensions then reduces to an $S^1$ dynamic in two dimensions, by  recognizing that the rotor precesses freely, leaving the polar mismatch angle $\delta\phi$ as the effective degree of freedom, and by recognizing that the field-parallel coupling in the reconnection region is strong enough to warrant a two-dimensional lattice in the field-perpendicular plane. This gives us the Adler equation, 
\begin{equation*} 
    \dot{\delta\phi_i} = \omega_i - \zeta \sin(\delta\phi_i),
\end{equation*}
and a Kuramoto model in the rotor's tangent plane (Eq.~\ref{eq:kuramoto}). A coupled $S^1$ oscillator lattice in two dimensions belongs to the $XY$ Universality class \cite{kosterlitz_ordering_1973}, or $O(2)$ symmetry group, an assessment that the simulations confirm via the evident hierarchical shock merger (Figure~\ref{fig:stats}a) and the universal scaling of the product $J_c\cdot T_e$ (Figure~\ref{fig:stats}c).

\subsection{Magnetic reconnection as a thermodynamic phase transition}


The foregoing compels a condensed matter treatment. The specific microscopic interaction details (such as exact particle trajectories governed by Lorentz forces) become irrelevant operators under renormalization flow (see, e.g., Ref.~\cite{wilson_renormalization_1971}), owing to a set of shared critical exponents at the critical point (see, e.g., Ref.~\cite{fisher_renormalization_1974}). In Appendix~B, we apply renormalization group theory (see Appendix~B) to Eq.~(\ref{eq:eom3d}), demonstrating that the system flows to a stable, scale-invariant fixed point. Consequently, the topology is agnostic to the specific interaction length $d_e$, which serves here merely as the necessary dimensional calibration for the universal geometry. The logistic breakdown of the ``frozen-in'' condition is therefore identified as the universal signature of the Berezinskii-Kosterlitz-Thouless (BKT) transition \cite{kosterlitz_ordering_1973,berezinskii_destruction_1971}, in which framework the 'critical current' $J_c$ represents the unbinding threshold for topological vortex pairs. Magnetic reconnection is thus identified as a thermodynamic phase transition.

The foregoing is supported in the recent literature. Kinetic simulations by Ref.~\cite{jara-almonte_thermodynamic_2021} have confirmed that the onset of collisionless reconnection exhibits the signature of a thermodynamic phase transition, characterized by a discontinuity in the current sheet's heat capacity. Our spintronic framework provides the topological mechanism explaining Ref.~\cite{jara-almonte_thermodynamic_2021}'s result: the `disordered phase' corresponds to the Adler-Ohmic slippage regime where gyrophase coherence is irreducibly lost.

Application of renormalization group theory (see Appendix~B) leads directly to our treatment of magnetic reconnection as phase slippage in an overdamped spintronic condensate. That we recover the conventional magnetohydrodynamic reconnection trigger (Eq.~\ref{eq:trigger0_m}), magnetic island coalescence (Figure~\ref{fig:stats}a), curvature drift (Eq.~\ref{eq:curvaturedrift}), and Bohm resistivity (Eq.~\ref{eq:etaspin_m}), empirically validates the description.

\subsection{Bohm resistivity \& collective transport}

The onset of magnetic reconnection in our overdamped spintronic condensate with increasing current $J$ (steepening magnetic field gradients), is shown in Figure~\ref{fig:stats}b) to be extremely sharp, modeled with a logistic function, consistent with observations of fast, ``bursty'', even transient magnetic reconnection \cite{frey_dayside_2019,southwoodWhatAreFlux1988,oksavikHighresolutionObservationsSmallscale2004,ivarsen_eastward_2025-1}.  This is consistent with our analysis of a magnetotail reconnection event observed by the spacecraft MMS-3, where the agyrotropy explodes vertically as the current builds up (Figure~\ref{fig:mms}). 

The observation of extremely sharp gyrophase bunching, quantified by a Von Mises concentration parameter saturating at $\Gamma>800$ in Figure~\ref{fig:mms}a), as well as the coherent red streak in Figure~\ref{fig:mms}b), challenges the traditional interpretation of anomalous resistivity as a stochastic or turbulent phenomenon. While kinematic turbulence or scattering would naturally increase entropy and smear the velocity distribution (leading to a low $\Gamma$), the data reveals an explosive transition into a singularly synchronized state where electron gyro-phases are strictly locked. This effective stiffness against thermal dispersion suggests the formation of a topologically locked condensate, where the electron fluid collectively maintains coherence to facilitate phase slippage, rather than interacting as a gas of independent particles.

This recognition constitutes a first-principles theoretical prediction and empirical verification of anomalous, or Bohm, resistivity ($\eta\propto T/B$), a long-time empirical assumption for space plasmas \cite{okudaInterpretationEnhancedDiffusion1972,millarIonCurrentsIonneutral1976,ottDiffusionStronglyCoupled2011}, whose onset as a topological phase transition here emerges naturally: the resistivity is proportional to the phase slip velocity, which saturates at the electron thermal speed $T_e$. The phase transition we observe is explosive, which matches the observed sudden release dynamics observed in solar flares \cite{heyvaertsEmergingFluxModel1977,tsunetaStructureDynamicsMagnetic1996}. 


The recovery of Bohm resistivity is a significant result. For some 70 years, the discrepancy between observed and theorized plasma diffusion rates, termed \emph{anomalous} diffusion, has been phenomenologically patched using Bohm resistivity \cite{hockneyComputerExperimentAnomalous1966,okudaInterpretationEnhancedDiffusion1972,okudaTheoryNumericalSimulation1973,millarIonCurrentsIonneutral1976,marchettiAnomalousDiffusionCharged1984,kaufmanExplanationBohmDiffusion1990,ottDiffusionStronglyCoupled2011,curreliCrossfieldDiffusionLowtemperature2014}. The overdamped spintronic condensate model of a magnetized plasma predicts that the onset of Bohm-like resistivity is an Adler-Ohmic bifurcation, suggesting that Bohm resistivity may be directly caused by electron gyrophase slippage, events that are already necessitated by the breakdown of the guiding center approximation \cite{scudder_fingerprints_2002,scudder_frozen_2015}. The natural consequence is that Bohm resistivity is a topological property of magnetized matter rather than a turbulent effect, arising at critical points such as magnetic reconnection, when the system effectively ignores microscopic restraints and yields to universal topological constraints.

In the next section, we describe the conditions of the plasma that the foregoing description is contingent on, applying renormalization group theory to demarcate the expected area of validity for our model.


\section{The overdamped spintronic condensate: insights from renormalization group flow} \label{sec:rg}

Our findings are contingent on the careful mapping of electron gyro-axes in a magnetized plasma to a 3D lattice of rotors aligned with the mean field via a Gilbert damping torque (Figure~\ref{fig:geometry}), essentially a \textit{classical, overdamped spintronic condensate}. Within this topological framework, the crossover scale $\xi_c$ emerges naturally as the electron inertial length $d_e$ (Eq.~\ref{eq:xi1}), physically anchoring the universality class. While the geometric mapping is demonstrably accurate, the validity of the overdamped condition deserves scrutiny.

Theoretically, this condition implies that the dissipative coupling dominates the inertial response at macroscopic scales. In terms of microphysics, this corresponds to a regime where electrons are subject to current-driven micro-instabilities (e.g., Buneman \cite{bunemanExcitationFieldAligned1963} or ion-acoustic modes \cite{li_role_2025}), triggered when the drift velocity exceeds a critical acoustic threshold ($v_d>C_s$). As detailed in Ref.~\cite{st.-mauriceTheoreticalFrameworkChanging2016}, increasing drift beyond this threshold generates intense turbulence, scattering and heating the electrons. This anomalous heating raises the stability threshold, creating a feedback loop where the drift is effectively clamped. Consequently, in the limit of strong driving, the system behaves as if governed by a drag force scaling with the thermal speed $v_\text{th}$ rather than Newtonian inertia. In this turbulent marginal stability regime, anomalous drag vastly exceeds the inertial term, rendering the electrons effectively overdamped.

Crucially, our renormalized group analysis (which we perform in Appendix~B) confirms that the inertial mass is an irrelevant operator that renormalizes to zero at macroscopic scales. Ref.~\cite{jafari_renormalization_2025} explicitly argues that the turbulent electric and magnetic fields inside the reconnection layer are Hölder singular (rough), causing bare gradients to be ill-defined. This necessitates treating the equations as an effective field theory with running parameters, justifying a renormalized mass $m_e(l)$ that flows to zero, $m_e(L) \to 0$, as the scale $L$ increases. This reduces the second-order Newtonian equation of motion to the first-order Adler equation (see Eqs.~\ref{eq:adler}, \ref{eq:plasmaadler}), effectively filtering out high-frequency cyclotron resonances to isolate the slow-manifold topological dynamics. Here, the inertia suppressed by renormalization is replaced by a thermodynamic closure, where work is dissipated directly into the thermal pool.

\subsection{Renormalization group stability of the inertial fixed point}

To formalize the foregoing, we shall, in this Section, perform a dynamic renormalization group analysis of a magnetized electron fluid in two spatial dimensions to determine the stability of the inertial (Hamiltonian) fixed point against dissipative perturbations. We construct a generic effective action containing operators for stiffness (exchange), inertia (time-reversal symmetric), and topological damping (time-reversal breaking). By performing a momentum-shell integration and rescaling space and time as $x' = b^{-1}x$ and $t' = b^{-z}t$, we calculate the scaling dimensions of the inertial and dissipative coupling constants. We demonstrate that at the inertial fixed point (dynamic critical exponent $z=1$), the dissipative operator is relevant, rendering the Hamiltonian description unstable at macroscopic scales. Conversely, at the dissipative fixed point ($z=2$), the inertial operator is irrelevant, scaling as $b^{-2}$. This establishes that the macroscopic dynamics of the critical state are universally governed by the overdamped fixed point, independent of the microscopic collision rate. Finally, we derive the crossover length scale $\xi_c$ separating the inertial and dissipative regimes.

We define the order parameter field $\mathbf{n}(\mathbf{x}, t)$ as the unit vector field representing the electron gyro-axis orientation. In the path integral formulation \cite{altland_condensed_2010}, we define the effective action $S_\text{eff}$,
\begin{multline} \label{eq:action}
    S_\text{eff} = \int d^d x dt \left[ \frac{I}{2} (\partial_t \mathbf{n})^2 + \right. \\ + \left. \zeta \mathbf{n} \cdot (\mathbf{n} \times \partial_t \mathbf{n}) - \frac{K}{2} (\nabla \mathbf{n})^2\right],
\end{multline}
where $I= N_e m_e$, $m_e$ being electron mass and $N_e$ the electron number density. $S_\text{inert} = (\partial_t \mathbf{n})^2/2$ is the inertial, kinetic energy of the rotor, and the dissipation term $S_\text{LL} = \zeta \mathbf{n} \cdot (\mathbf{n} \times \partial_t \mathbf{n})$ is the Landau-Lifshitz damping \cite{landau_theory_1935,gilbert_phenomenological_2004}. Stiffness $S_\text{stiff}=K(\nabla \mathbf{n})^2/2$ is the energy cost of spatial gradients in the magnetic topology, i.e., magnetic tension \cite{auerbach_interacting_2012}, with $K=B^2/\mu_0$. $d$ is the number of dimensions, and, in what follows, we consider the current sheet plane in two dimensions ($d=2$).

\subsection{Scaling Dimensions and Fixed Point Stability}

We next perform a Kadanoff block-spin transformation \cite{kadanoff_scaling_1966,wilson_renormalization_1971}, integrating out short-wavelength fluctuations, rescaling the system to the macroscopic reconnection sheet size. We introduce a scaling factor $b > 1$, which yields spatial rescaling, $\mathbf{x}' = b^{-1} \mathbf{x}$, and temporal rescaling, $t' = b^{-z} t$. Here, $z$ is the dynamic critical exponent, which determines the relative scaling of space and time, via the dispersion relation $\omega \sim k^z$. Balancing the inertial term $I(\partial_t n)^2$ against the stiffness $K(\nabla n)^2$ in Fourier space yields $I \omega^2 \sim K k^2$, so that
\begin{equation}
    \omega \sim \sqrt{\frac{K}{I}} k.
\end{equation}
Balancing the damping term $\zeta n (n \times \partial_t n)$ against stiffness yields $\zeta \omega \sim K k^2$ (where drop vector factors for brevity). We have,
\begin{equation}
    \omega \sim \frac{K}{\zeta} k^2.
\end{equation}
Since frequency scales quadratically with wavenumber, time scales as the \textit{square} of space ($t \sim x^2$), corresponding to $z=2$, consistent with Model A/B dynamics in the Halperin-Hohenberg classification \cite{hohenberg_theory_1977}.

The stiffness term in Eq.~(\ref{eq:action}) is next chosen conveniently as the invariant reference to fix the scaling dimension of the field $\mathbf{n}$. For that term,
\begin{equation}
    \int d^d x (\nabla \mathbf{n})^2 \to b^d \cdot b^{-2} \int d^d x' (\nabla' \mathbf{n})^2 = b^{d-2} S_\text{stiff}.
\end{equation}

To determine which physics dominates at the macroscopic scale, we analyze the scaling dimensions of the coupling constants for inertia ($m(b)$) and dissipation ($u_\text{LL}$) relative to the stiffness term ($K$). At the fixed point (where stiffness is scale-invariant), we observe how the inertial and damping terms flow as the rescaling factor $b \to \infty$. We have,
\begin{equation}
    x' = b^{-1}x \;\;\;\text{ and }\;\;\; t' = b^{-z}t,
\end{equation}
which we apply to the three terms in the action integral Eq.~(\ref{eq:action}),
\begin{align}
  \int d^2x dt (\partial_t \mathbf{n})^2 \sim b^2 \cdot b^z \cdot (b^{-z})^2 &= b^{2-z}, \\
    \int d^2x dt (\mathbf{n} \times \partial_t \mathbf{n}) \sim b^2 \cdot b^z \cdot (b^{-z}) &= b^2, \\
  \int d^2x dt (\nabla \mathbf{n})^2 \sim b^2 \cdot b^z \cdot (b^{-1})^2 &= b^z .
\end{align}

We define the renormalized dimensionless couplings $u$ by dividing the inertial and dissipative scales by the stiffness scale ($b^z$),
\begin{align}
    m(b)' & \propto b^{2-2z} m(b), \\   
    u_\text{LL}' & \propto b^{2-z} u_\text{LL}.
\end{align}
which we use to test the two physical regimes, discriminating stable from unstable points (see, e.g., Ref.~\cite{hohenberg_theory_1977}).

The stability of a fixed point is determined by how small perturbations to the action evolve under the renormalization group flow. If a coupling constant $u$ scales as $b^y$ with $y > 0$ (a relevant operator), the perturbation grows macroscopically as we coarse-grain to larger scales ($b \to \infty$), driving the system away from the fixed point. This implies the fixed point is unstable. Conversely, if $u$ scales as $b^y$ with $y < 0$ (an irrelevant operator), the perturbation diminishes at large scales, implying the fixed point is stable (an attractor).

\textit{Hypothesis A:} setting $z=1$, we observe that $m(b)\propto b^0$ and $u_\text{LL}\propto b^1$, meaning that the former is marginal and stable, and the latter is relevant. At this point, the dissipative term grows linearly with scale, making it unstable; any non-zero dissipation (resistivity and damping) will eventually dominate the dynamics at large scales, a known feature of quantum critical points in dissipative environments \cite{chakravarty_two-dimensional_1989,sachdev_quantum_1999}. This is the inertial fixed point, since $z=1$ implies dispersion $\omega\sim k$, characteristic of antiferromagnetic spin waves or undamped phonons \cite{auerbach_interacting_2012}. The inertial point being unstable is consistent with the turbulent fields near the reconnection being rough, mandating renormalization \cite{jafari_renormalization_2025}.

\textit{Hypothesis B:} setting $z=2$, we observe that $m(b)\propto b^{-2}$ and $u_\text{LL}\propto b^0$; the former is irrelevant and the latter is marginal and stable. This is the dissipative, Adler-Ohmic fixed point ($\omega\sim k^2$) and we note that the inertial term decays quadratically with scale, making the overdamped fixed point stable. 

The UV limit at microscopic scale, for small $b$, electron inertia (the Lorentz force) may be significant: $z \approx 1$. However, because the dissipative operator is relevant at the inertial fixed point, the system flows inevitably toward the $z=2$ fixed point. At the macroscopic scale of the reconnection site ($b \to \infty$), the inertial Lorentz force is an irrelevant operator; it scales as $b^{-2}$). The collective behavior at the macroscopic reconnection site (the IR limit) is therefore strictly governed by the topological dissipation term.

\begin{figure}
    \centering
    \includegraphics[width=0.34\textwidth]{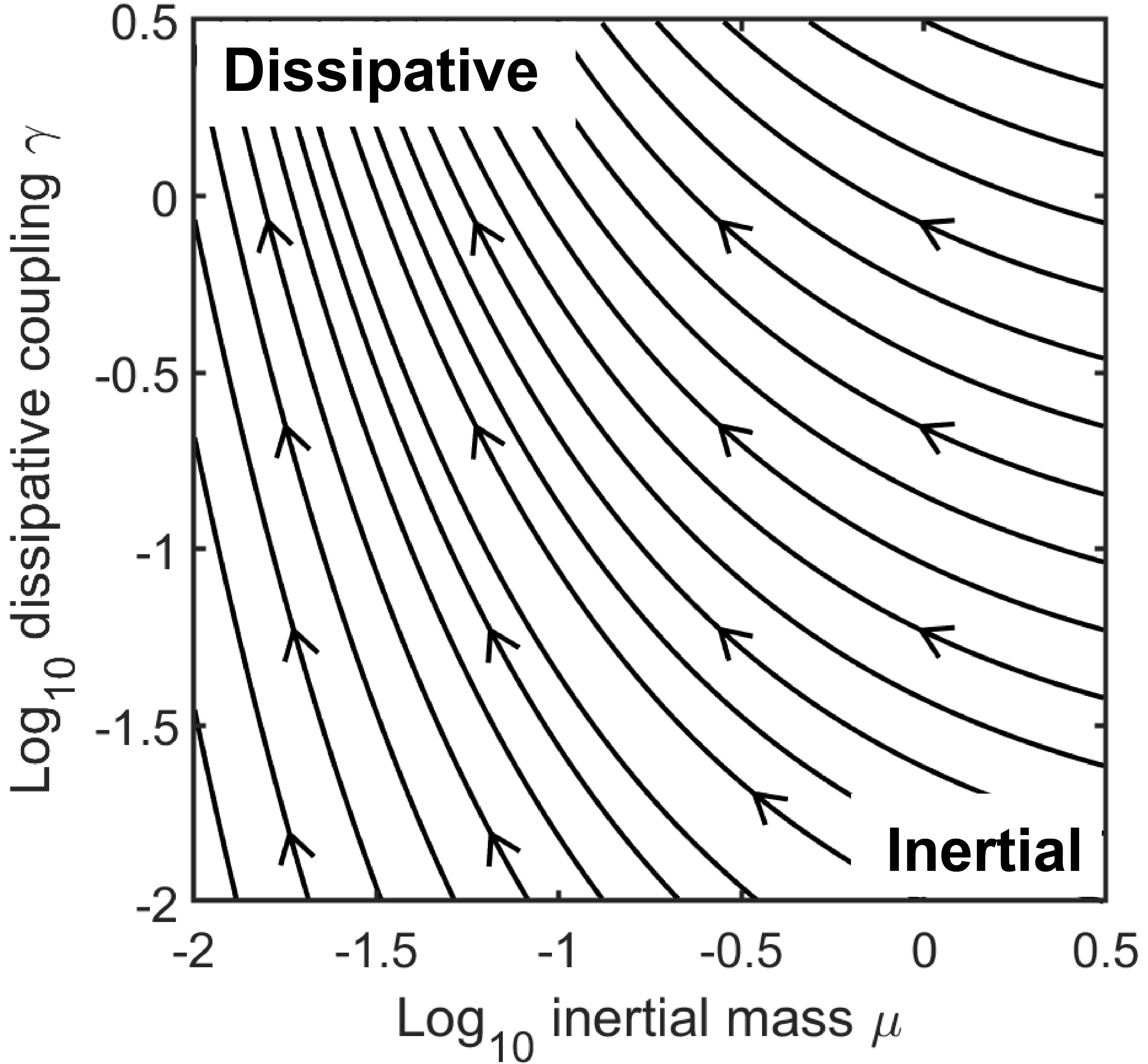}
    \caption{
    \textbf{Renormalization Group (RG) flow diagram illustrating the universality of the overdamped limit.} Horizontal axis represents inertial mass $m$ (Lorentz force coupling strength), and the vertical axis represents dissipative coupling $u_\text{LL}$ (the friction from magnetic topology). Streamlines visualize the trajectory of the system parameters as the scale increases $(b\to\infty)$.}
    \label{fig:rg}
\end{figure}

We note that, the inertial fixed point being unstable, any non-zero dissipation at the micro-scale will cause the system to flow to the dissipative fixed point, making the overdamped condition the attractor of the plasma undergoing magnetic reconnection (see Figure~\ref{fig:rg} for a flow diagram based on the scaling relations). The dissipative fixed point is inertia-less, leading directly and naturally to the Adler equation.

\subsection{The crossover scale}

We determine the crossover length scale $\xi_c$ by finding the wavenumber $k_c$ where the inertial and dissipative contributions to the action are of equal magnitude. From setting the Fourier space representation of the operators in Eq.~(\ref{eq:action}) equal to zero, we obtain the dispersion relation,
\begin{equation}
N_em_e\omega^2 + i\zeta\omega - \frac{B^2}{\mu_0}k^2 = 0.
\end{equation}
The transition occurs when the inertial term matches the damping term, $\omega_c = \zeta/N_em_e $. We substitute this into the inertial (antiferromagnetic) dispersion relation $\omega_c = v_{sw} k_c$, where the inertial spin wave velocity is given by,
\begin{equation}
v_{sw} = \sqrt{\frac{K}{I}} = \frac{B}{\sqrt{\mu_0N_em_e}} = v_{A},
\end{equation}
identifying the electron Alfv\'{e}n speed $v_A$. Solving for $k_c$, we find that
\begin{equation}
\frac{\zeta}{N_em_e} = v_A k_c,
\end{equation}
which, solving for the length scale $\xi_c = 1/k_c$, yields
\begin{equation} \label{eq:xi1}
\xi_c = \frac{N_em_ev_A}{\zeta}.
\end{equation}
To determine $\xi_c$, we strictly evaluate the damping coefficient $\zeta$ by analyzing the asymptotic limits of the effective action.
In Section~IV-B, we identified that the Goldstone mode of the precessing gyro-axes supports spin waves (magnons) that share the quadratic dispersion relation of whistler waves \cite{gordeev_electron_1994,volovik_universe_2003}. The topological model itself is scale-invariant, and so we must match the $z=2$ fixed point to Hall-magnetohydrodynamics (whistler mode), which identifies the coupling $\zeta$ as the gyroscopic density, $\zeta=N_eeB$. Eq.~(\ref{eq:xi1}) then yields $\xi_c=v_A/\Omega_{ce}$. Using the definition of the Alfv\'{e}n speed $v_A = c\Omega_{ce}/\omega_{pe} $ (with $\omega_{pe}$ being the electron plasma frequency), we recover the fundamental scale:
\begin{equation}
\xi_c = d_e.
\end{equation}
This is exactly the electron inertial length. The RG flow thereby predicts a universal transition from ballistic (inertial) to overdamped (diffusive) dynamics at scales larger than $d_e$.

\section{Discussion~B}

\textit{A posteriori} support for the foregoing macroscopic description comes from considering that the overdamped condition directly leads to a slippage electric field $\mathbf{E}_\text{eff}$ that scales identically to the current tension electric field $\mathbf{E}_{CT}$ (Eq.~\ref{eq:ect}) \cite{luo_current_2024}, caused by electron inertia (see Appendix~A).

The fact that the resulting gyro-axis slip field mirrors the inertial current tension field, as well as the recovery of both the ideal magnetohydrodynamic breakdown threshold (Eq.~\ref{eq:trigger0_m}) and curvature drift (Eq.~\ref{eq:curvaturedrift}), implies that phase stiffness is a topological equivalence for electron inertia (see, e.g., Refs.~\cite{ivarsen_information_2025,ivarsen_onsager_2025-1}). Our model thereby successfully isolates the universal energetic barrier for reconnection, treating the kinetic resistance to turning as a mathematically equivalent thermodynamic potential.

\subsection{Area of validity}

The general validity of our result is contingent on the original mathematical abstraction involved in modeling electron gyro axes as three dimensional rotors that align on the unit surface $S^2$ with the local mean magnetic field. This alignment leads to precession around the mean-field and projects conservative dynamics to phase alignment ($S^1$, see Figure~\ref{fig:geometry}a ,b) the tilted washboard potential, and the Adler equation (Eq.~\ref{eq:adler}). This result, and the subsequent $XY$-model simulations (Figure~\ref{fig:stats}), leads directly to magnetic reconnection triggering (Eq.~\ref{eq:trigger0_m}) Bohm resistivity scaling $\eta_0\propto B^{-1}$, thereby providing a first-principles solution to the longest-standing problem in plasma physics.

The foregoing mathematical abstraction necessitates renormalization group theory to justify the lack of inertia in the electrons (Section~\ref{sec:rg}), and our model is therefore a statistical mechanical description of magnetic reconnection. The overdamped conditions (which recovers the Landau-Lifshitz-Gilbert equation) is supported by the \textit{in-situ} observations by MMS (Figures~\ref{fig:mms0}, \ref{fig:mosaic2}, and \ref{fig:mms}). The analysis demonstrates that Ref.~\cite{torbert_electron-scale_2018}'s `crescents' become a funnel of suprathermal electrons that collectively move, or slip, across field-lines. The singular (delta-function) phase space confinement implies that the electron fluid has undergone a phase transition into a topologically locked condensate, in clear validation of our initial mathematical abstraction. This field-perpendicular \textit{slip} ensures that the two-dimensional mapping in Figure~\ref{fig:geometry}b) is justified (at which point the model recovers the Kuramoto model under renormalization group flow, see Figures~\ref{fig:geometry} and \ref{fig:rg}). 
Our model is therefore well-supported by a key \emph{in-situ} observation of magnetic reconnection.

The novel mathematical bridge deserves close scrutiny. One such avenue of falsification comes through the consideration the the effortless precession around the mean magnetic field of the electron gyro axes, which drives a Goldstone mode quasiparticle, the ferromagnetic magnon.

\subsection{Outlook: Whistler magnons}


%

To describe the spin waves that our model predicts, we solve for the dispersion relation and perform the mapping ($\xi\to d_e$ and $\omega_i\to \Omega_{ce}+\delta \omega_i$, see Appendix~A). In so doing, we recover in the condensate the presence of ferromagnetic magnons, spin waves that ripple through the precessing, slipping electron gyro-axes,  which advect enstropy (see, e.g., Ref.~\cite{chumak_magnon_2015} for a review). The dispersion relation reads (see Appendix~B), 
\begin{equation}\label{eq:dispersion1}
\omega(k) = \Omega_{ce} d_e^2 k^2 - i \zeta d_e^2 k^2,
\end{equation}
which is the exact dispersion relation for whistler-mode plasma waves \cite{treumann_advanced_1997,burch_electron-scale_2016}. Since the gyro-axes and the magnetic topology are mutually dependent, this results indicates that, in the phase slipping regime, magnons propagate through the precessing gyro-axes in tandem with whistler plasma waves, the latter of which then obtain a \textit{topological footprint} in the slipping gyro-axes. $\zeta$ in Eq.~(\ref{eq:dispersion}) follows from the Landau-Lifshitz torque (Gilbert damping, Eq.~\ref{eq:llb_m}). In a collisionless plasma, this energy loss corresponds to Landau damping or wave-particle resonance extracting energy from the topological rearrangement, linking enstropy production in our model to heating in the plasma. This implicates the interoperability of coupled topological and kinetic simulations of magnetic reconnection. 

Our model is scale-invariant, and so the mean-field coupling $\zeta$ can be predicted in observations of whistler wave-particle interactions. By capturing the conventional whistler-mode energy transport with an equivalent transport of enstrophy, the magnons should, in principle, be visible in electron phase density observations of \emph{in-situ} whistler wave propagation. 


\section{Conclusion}


We have formulated the Landau-Lifshitz-Gilbert equation for an overdamped spintronic condensate, and demonstrate that this model makes unique predictions for magnetic reconnection, namely the onset of anomalous resistivity. This justifies the application of renormalized group theory (Appendix~B), which, in turn, indicates that magnetic reconnection can be understood as a topological phase transition of the $XY$ universality class \cite{kosterlitz_ordering_1973}. We provide empirical evidence that the onset of electron gyrophase slippage follows universal BKT scaling laws \cite{berezinskii_destruction_1971},  and we demonstrate that the model is consistent with observations of magnetotail reconnection by the MMS mission.  The model predicts a non-linear anomalous resistivity that rigorously recovers the empirical Bohm scaling. We thereby provide a first-principles origin for anomalous diffusion in space plasmas.

Our results are consistent with recent field-theoretical treatments of magnetohydrodynamics  under renormalization group flow \cite{jafari_renormalization_2025}, and we find that the description of guiding centre approximation breakdown under RG flow anchors magnetic reconnection to the same fundamental dynamics that govern ferromagnetism: the Landau-Lifshitz-Gilbert equation \cite{lakshmanan_fascinating_2011}. 
Just as Ginzburg-Landau theory in condensed matter physics abstracts the complex electrodynamics of electron cooper pairs in superconductors \cite{gorkov_microscopic_1959}, our Adler-Ohmic model abstracts the chaotic particle kinetics during strong electron agyrotropy to isolate the collective, anomalous transport involved in the critical trigger-point of magnetic reconnection. 

\section*{Acknowledgements}
This work is supported by the European Space Agency’s Living Planet Grant No. 1000012348. The author is grateful to Y. Miyashita, O. Nestande, D. Knudsen, PT. Jayachandran, and K. Douch for stimulating discussions. Google's Gemini 3.0 Pro has been used to assist mathematical formalism and coding in \textsc{matlab}.


\section*{Appendix A: Theoretical foundation}

Following Ref.~\cite{ivarsen_information_2025}, we start with the Adler equation (Eq.~\ref{eq:adler}). We calculate the time required for a complete $2\pi$ phase slip:
\begin{equation} 
    T_\text{slip} = \int_0^{2\pi}\;\frac{d\delta\phi}{\dot{\delta\phi}} = \int_0^{2\pi}\frac{d\delta\phi}{\omega_i - \zeta \sin(\delta\phi)} = \frac{2\pi}{\sqrt{\omega_i^2-\zeta^2}}, 
\end{equation} 
where we assumed the running state ($\omega_i > \zeta$). Consequently, the slip velocity is defined as the inverse of the slip period:
\begin{equation} \label{eq:vslip2} 
    v_\text{slip} \equiv \langle \dot{\delta\phi}\rangle = \frac{2\pi}{T_\text{slip}} = \sqrt{\omega_i^2-\zeta^2},
\end{equation}
where we note that $v_\text{slip}$ is a time-averaged quantity (over slip cycles), and where we stress that this is the slippage of the phase of the \textit{gyro-axis}, not the electron's gyrophase, and that the latter slips when the former does so. Magnetic field-reversal follows from the gyro-axis slipping $\pi$.

\subsection{The Landau-Lifshitz-Gilbert equation}

Examining the hydrodynamic limit, we must assume that intrinsic chirality is aligned with the mean field, $\hat{\omega}_i\approx\omega_i\hat{\Psi}$. We can then write Eq.~(\ref{eq:eom3d}) as,
\begin{equation} \label{eq:simpl}
    \frac{\partial\hat{n}}{\partial t} \approx \omega_i (\hat{\Psi} \times \hat{n}) + \zeta [\hat{n} \times (\hat{\Psi} \times \hat{n})].
\end{equation}
Next, we perform a Taylor expansion of the local field around the chiral rotor. This introduces the Laplacian, transforming the discrete lattice into a continuous medium \cite{ivarsen_information_2025},
\begin{equation} \label{eq:continuum}
    \hat{\Psi}(\mathbf{r}, t) \approx \rho \left( \hat{n} + \xi^2 \nabla^2 \hat{n} \right),
\end{equation}
where $\xi$ is the correlation length (interaction radius), and $\nabla^2 \hat{n}$ is the curvature of the field. We substitute Eq.~(\ref{eq:continuum}) into Eq.~(\ref{eq:simpl}),
\begin{equation} \label{eq:simpl2}
    \frac{\partial\hat{n}}{\partial t} \approx -\omega_i \xi^2 (\hat{n} \times \nabla^2 \hat{n}) +\zeta \xi^2 [\hat{n} \times (\hat{n} \times \nabla^2 \hat{n})].
\end{equation}
Setting $\boldsymbol{H}_\text{eff}=\nabla^2\hat{n}$, we recover the Landau-Lifshitz-Gilbert equation in standard form \cite{lakshmanan_fascinating_2011},
\begin{equation}\label{eq:llb0}
    \frac{\partial \hat{n}}{\partial t} = -\frac{\gamma}{1+\alpha^2} (\hat{n} \times \mathbf{H}_{\text{eff}}) - \frac{\gamma \alpha}{1+\alpha^2} \hat{n} \times (\hat{n} \times \mathbf{H}_{\text{eff}}).
\end{equation}
Matching the terms yields the constitutive relations,
\begin{equation}
    \alpha = \frac{\zeta}{\omega_i},
\end{equation}
and
\begin{equation} 
\gamma = \omega_{i} \xi^2 (1 + \alpha^2),
\end{equation}
allowing us to write the Landau-Lifshitz-Gilbert equation on the form,
\begin{equation}\label{eq:llb}
    \frac{\partial \hat{n}_i}{\partial t} = -\omega_{i} \xi^2 (\hat{n}_i \times \mathbf{H}_{\text{eff}}) - \zeta \xi^2 \hat{n}_i \times (\hat{n}_i \times \mathbf{H}_{\text{eff}}).
\end{equation}
In the next section, we shall use this model to rigorously describe the phase mismatch between the electron gyro-axis and the local mean field in a magnetized plasma.

\subsection{Gyrophase slippage \& spintronic reconnection}

Consider the angular motion of the electron gyro-axis itself as a source of \textit{chirality}, or frustration, $\omega_i$, a distribution that obeys $\langle \omega_i\rangle = \Omega_{ce}$, the electron cyclotron frequency. Then, an increase in the spread of $\omega_i$, which correspond to thermal broadening and relativistic mass correction, will proportionally increase the sum total enstropy, eventually reaching the point when the magnetic topology can no longer support the guiding centre approximation, at which point magnetic reconnection will have occurred. The trigger is an Adler-Ohmic crossover (Eq.~\ref{eq:adler}),  and after this, continuous gyrophase slippage causes an irreversible reduction in energy, by inducing chaotic electron orbits \cite{speiser_particle_1991}.

To implement the foregoing, we apply the minimal model that was defined in the previous section. The vector $\hat{n}_i$ becomes the electron's magnetic moment $\boldsymbol{\mu}_i$ aligning against $\mathbf{B}$, but whose axis of precession in general takes the form $\hat{n}_i$. The force that seeks to stabilize the orbit is the dissipative Landau-Lifshitz torque, which acts to align the electron's gyration axis  with the local magnetic field \cite{landau_collected_1965}. This alignment torque takes the Gilbert form:
\begin{equation} 
\boldsymbol{\mathcal{T}}_\text{LL} = \zeta \;\hat{n}_i\times (\boldsymbol{b}_i \times \hat{n}_i),
\end{equation}
where $\boldsymbol{b}=\mathbf{B}/B$ is the unit magnetic field vector. When $\hat{n}_i$ and $\boldsymbol{b}$ aligns, the electron gyro-axis is perfectly ``frozen-in,'' and, as we shall demonstrate, \textit{trapped}. The mismatch between the two, $\delta\phi$, therefore measures whether the electron is orbiting a magnetic field-line in a stable manner or entering a chaotic orbit \cite{speiser_particle_1991}.

We can now write the plasma Adler equation,
\begin{equation}
\label{eq:plasmaadler} \dot{\delta\phi} = v_{\text{th},i} \left( \frac{1}{L_{\nabla B}} - \frac{1}{\rho_L}\sin(\delta\phi) \right), \end{equation}
where $\Omega_{ce}=v_{\text{th},i}/\rho_L$ is the electron cyclotron frequency, $\rho_L$ being the Larmor radius of the precessing electron  \cite{stasiewicz_finite_1993}, and where we replace chirality $\omega_i$ with the rate at which the magnetic field direction changes, $v_{\text{th},i}/L_{\nabla B}$, where $L_{\nabla B}$ is the length-scale of the magnetic gradient and $v_{\text{th},i}$ is the (thermal) speed of the electron.

Eq.~(\ref{eq:plasmaadler}) expresses that electrons are `running' when they can no longer sustain perfect orbits around a field line, crossing the limit of adiabatic invariance \cite{buchner_regular_1989}. Here, we recover the magnetohydrodynamic threshold for magnetic reconnection,
\begin{equation} \label{eq:trigger0}
    \rho_L > L_{\nabla B}.
\end{equation}
This threshold is conventionally derived through considering the magnetic moment $\boldsymbol{\mu}$ as no longer conserved (the breakdown of adiabatic invariance). In the present article, we have derived the same result purely from a consideration of vector alignment on the unit sphere.

Next, we shall demonstrate that the inherently non-linear and explosive nature of the Adler slip velocity (Eq.~\ref{eq:vslip2}) fuels the anomalous resistivity that accompanies magnetic reconnection.

\subsection{Magnetic reconnection \& anomalous resistivity}

To quantify the anomalous resistivity, we decompose the electron drift velocity,
\begin{equation} 
\mathbf{v}_\text{total} = \mathbf{v}_\text{ideal} + \mathbf{v}_\text{slip}, 
\end{equation} 
where $\mathbf{v}_\text{ideal}=(\mathbf{E}\times\mathbf{B})/B^2$ is the ideal, frozen-in drift. The deviation from this ideal state is the spintronic slip velocity, whose magnitude is governed by the Adler-Ohmic bifurcation derived in Sec.~\label{sec:intro}: 
\begin{equation} 
|\mathbf{v}_\text{slip}| = \rho_L \sqrt{\frac{v_{\text{th},i}^2}{L_{\nabla B}^2} - \Omega_{ce}^2}, 
\end{equation} 
where we multiplied the angular slip velocity (Eq.~\ref{eq:vslip2}) with the Larmor radius $\rho_L$. This slip motion generates an anomalous motional electric field (resistivity), \begin{equation} \label{eq:eeff}
\mathbf{E}_\text{eff} = -\mathbf{v}_\text{slip} \times \mathbf{B}, \end{equation} 
which breaks the frozen-flux condition whenever $\rho_L > L_{\nabla B}$. In the running state ($\omega_i\gg\zeta$), we can write,
\begin{equation} \label{eq:curvaturedrift}
    |\mathbf{v}_\text{slip}| \approx \frac{v_{\text{th},i} \rho_L}{L_{\nabla B}},
\end{equation}
where we observe that the magnitude of the cross-field slippage scales identically to the standard curvature drift \cite{northrop_adiabatic_1963}, yet represents a diffusive flux perpendicular to the magnetic field. The magnitude of Eq.~(\ref{eq:eeff}) is given by, 
\begin{equation}
E_\text{eff} = |\mathbf{v}_\text{slip}| B,
\end{equation}
and after inserting the non-linear Adler slip velocity, we obtain,
\begin{equation}\label{eq:driveform}
E_\text{eff} \approx B  \rho_L \sqrt{\left(\frac{v_{\text{th},i}}{L_{\nabla B}}\right)^2 - \Omega_{ce}^2}.
\end{equation}
To find the resistivity $\eta$, we must express the field gradient scale $L_{\nabla B}$ in terms of the current density $J$, which, using Amp\`{e}re's law ($\nabla \times \mathbf{B} = \mu_0 \mathbf{J}$, $\mu_0$ being magnetic permittivity), leads to the following proximation,
\begin{equation}
\frac{1}{L_{\nabla B}} \approx \frac{|\nabla B|}{B} \approx \frac{\mu_0 J}{B},
\end{equation}
which we can readily substitute into Eq.~(\ref{eq:driveform}),
\begin{equation}
E_\text{eff} \approx B \rho_L \sqrt{ \left( v_{\text{th},i} \frac{\mu_0 J}{B} \right)^2 - \Omega_{ce}^2 }.
\end{equation}
From this, and Eq.~(\ref{eq:trigger0}), we observe the threshold spintronic critical current $J_c$:
\begin{equation}
v_{\text{th},i} \frac{\mu_0 J_c}{B} = \Omega_{ce} \implies J_c = \frac{e B^2}{m \mu_0 v_{\text{th},i}},
\end{equation}
or,
\begin{equation}
    J_c = \frac{B}{\mu_0\rho_L},
\end{equation}
which corresponds to the current density where the electron Larmor radius equals the current sheet width, $\rho_L = L$).

To transform this into the standard Ohmic form $E = \eta J$, we factor out the drive coefficient $\left( v_{\text{th},i} \mu_0/B \right)^2$ from the square root. This cancels the magnetic field $B$, leaving the current $J$ as the primary variable,
\begin{equation}
E_\text{eff} \approx \left( \mu_0 \rho_L v_{\text{th},i} \right) J \sqrt{ 1 - \left( \frac{J_c}{J} \right)^2 },
\end{equation}
revealing the spintronic anomalous resistivity as the coefficient,
\begin{equation} \label{eq:etaspin}
\eta_{spin}(J) = \eta_0 \sqrt{1 - \left(\frac{J_c}{J}\right)^2} \quad \text{for } J > J_c.
\end{equation}
Here, we make two observations:

\textit{(1)} the characteristic resistivity scale is $\eta_0 = \mu_0 \rho_L v_{\text{th},i}$,
which, since $\rho_L v_{\text{th},i} \approx k_B T/e B$, means that we recover the Bohm resistivity scaling \cite{braginskiiTransportProcessesPlasma1965,ottDiffusionStronglyCoupled2011}), a standard phenomenological assumption in space plasmas \cite{treumann_advanced_1997,burch_electron-scale_2016}. 

\textit{(2)} the spintronic critical current provides a hard trigger, $\sqrt{1 - (J_c/J)^2}$, below which the anomalous resistivity is strictly zero (ideal magnetohydrodynamics). when the current density $J$ exceeds the critical threshold $J_c$ (the Adler limit), the electrons cascade non-linearly into a resistive state, triggering magnetic reconnection.

\subsection{Spin waves}

The precession of the gyro-axis around the mean field, which is, under the guiding centre approximation, assumed to be effectively averaged out, can, during gyro-axes slippage, drive a Goldstone mode, manifesting as ferromagnetic magnons that propagate through the phase-slipping rotors. We can derive the dispersion relation from Eq.~(\ref{eq:llb}), which yields (see Appendix~B),
\begin{equation}\label{eq:dispersion}
    \omega(k) = \frac{\gamma k^2}{1 + \alpha^2} - i \frac{\gamma \alpha k^2}{1 + \alpha^2},
\end{equation}
where the real part corresponds to the magnon and the imaginary part as its decay. 

We now apply the specific plasma parameters: 

\textit{(1)} $\omega_i \approx \Omega_{ce}$, which equates chirality with the electron cyclotron frequency, while recognizing that it should strictly read $\omega_i\to\Omega_{ce}+\delta\omega$, where $\delta\omega$ averages out.

\textit{(2)} $\xi \approx d_e$ (setting the correlation length approximately equal to the electron inertial length).

The final derived coefficient for the spintronic stiffness is,
\begin{equation}
\gamma = \Omega_{ce} d_e^2 (1 + \alpha^2),
\end{equation}
allowing us to write the dispersion relation as,
\begin{equation}\label{eq:dispersion1}
\omega(k) = \Omega_{ce} d_e^2 k^2 - i \zeta d_e^2 k^2.
\end{equation}

To see the significance of this result, consider the following. A magnon ripples through the phase-slipping rotors. The rotors, in turn, being electron gyro-axes, drag the magnetic topology $\hat{b}$, meaning that the magnons should be visible in  electric and magnetic fields measured \textit{in-situ}. This is because the magnons experience friction ($\alpha$ in Eq.~\ref{eq:llb}), by which the spin wave's coherence is thermalized to the dissipative bath. Eq.~(\ref{eq:dispersion1}) describes the dispersion relation for whistler-mode plasma waves, which are excited in the same region, after which they propagate outward, damping their energy into the electrons and thereby heating them \cite{burch_electron-scale_2016,cao_mms_2017}. The ferromagnetic magnons that propagate through the slipping gyro rotors in the spintronic condensate are therefore the topological footprint of whistler waves.

\subsection{Verification}

The foregoing effective electric field $\mathbf{E}_\text{eff}$ is consistent with recent research that considered the effects of the current tension electric field \cite{luo_current_2024},
\begin{equation} \label{eq:ect}
    \mathbf{E}_{CT} = -\frac{m}{e} \left[ \left( \frac{\mathbf{J}}{en} \cdot \nabla \right) \frac{\mathbf{J}}{en} \right],
\end{equation}
that is, the electron inertia associated with the spatial variation of the current flow. $E_{CT}$ is often ignored as a small number, but proven consequential inside the electron diffusion region \cite{luo_current_2024}, contributing to the reconnection electric field. In our framework, $E_{CT}$ appears when the current tension can no longer sustain electron gyrations, causing the resistive drifts (Eq.~\ref{eq:etaspin}). In Figure~\ref{fig:geometry}c we illustrate this term (in blue), along with the reconnection X-line, with aligned gyro-axes in black compasses, and gyrophase slipping electrons in red compasses.

It is crucial to distinguish the physical nature of the spintronic slip velocity, $\mathbf{v}_{slip}$, from the standard guiding center drifts of ideal magnetohydrodynamics. While the magnitude of the slip velocity in the running state, $|\mathbf{v}_{slip}| \approx v_{\text{th},i}\rho_L/L_{\nabla B}$, algebraically recovers the scaling of curvature drift, the two represent topologically distinct regimes of motion. Standard curvature drift is an adiabatic process: the electron's gyrophase orbit remains closed and the first adiabatic invariant (magnetic moment $\mu$) is conserved. In contrast, the Adler-Ohmic slip represents a dissipative failure of this orbit closure. When the drive exceeds the lock ($\omega_i > \zeta$), the electron crosses the limit of adiabatic invariance, and the gyration effectively 'opens' into a chaotic spiral. Unlike adiabatic drift, which is reversible and performs no work, this phase slippage generates entropy, manifesting macroscopically as the anomalous resistive field $\mathbf{E}_\text{eff}$ that breaks the frozen-in topology. 

\section*{Appendix C: the dispersion relation}

We begin with the Landau-Lifshitz-Gilbert equation on the form,
\begin{equation} \label{eq:disp1}
    \frac{\partial \hat{n}}{\partial t} = -\gamma \hat{n} \times H_\text{eff} + \alpha \hat{n} \times \frac{\partial \hat{n}}{\partial t}
\end{equation}

We consider the polarized case, and introduce a small transverse perturbation $\vec{m}$ (the magnon),
\begin{equation}
    \hat{n}(\vec{r}, t) \approx \hat{z} + \vec{m}(\vec{r}, t),
\end{equation}
where $\vec{m} = (m_x, m_y, 0)$ and $|\vec{m}| \ll 1$. We substitute this into the effective field definition $H_\text{eff} \approx \nabla^2 \vec{m}$, and then substitute this into Eq.~(\ref{eq:disp1}), keeping only linear terms,
\begin{equation}
    \frac{\partial \vec{m}}{\partial t} = -\gamma (\hat{z} \times \nabla^2 \vec{m}) + \alpha (\hat{z} \times \frac{\partial \vec{m}}{\partial t}).
\end{equation}
We solve this for the eigenmodes by assuming a plane wave solution with frequency $\omega$ and wavenumber $k$:
\begin{equation}
    \vec{m}(\vec{r}, t) = \vec{m}_0 e^{i(\vec{k} \cdot \vec{r} - \omega t)}.
\end{equation}
Operators transform as $\partial/\partial t \to -i\omega$ and $\nabla^2 \to -k^2$, which yields,
\begin{equation}
    -i\omega \vec{m} = -\gamma [\hat{z} \times (-k^2 \vec{m})] + \alpha [\hat{z} \times (-i\omega \vec{m})].
\end{equation}
We simplify the cross products (since $\hat{z}$ points upwards),
\begin{equation}
    -i\omega \vec{m} = \gamma k^2 (\hat{z} \times \vec{m}) - i\alpha\omega (\hat{z} \times \vec{m}),
\end{equation}
or,
\begin{equation}
    -i\omega \vec{m} = (\gamma k^2 - i\alpha\omega) (\hat{z} \times \vec{m}).
\end{equation}
To solve for $\omega$, we apply the circular polarization ansatz naturally favored by chiral systems: Let $m_+ = m_x + i m_y$. The cross product operation $\hat{z} \times \vec{m}$ corresponds to multiplication by $-i$:
\begin{equation}
    (\hat{z} \times \vec{m})_+ = -i m_+.
\end{equation}
Substituting this scalar equivalent into the vector equation:
\begin{equation}
    -i\omega m_+ = (-i\gamma k^2 - \alpha\omega) m_+,
\end{equation}
which, after dividing by the amplitude $m_+$ (assuming non-zero perturbation), yields,
\begin{equation}
    -i\omega = -i\gamma k^2 - \alpha\omega.
\end{equation}
We rearrange this to isolate $\omega$,
\begin{equation}
    \omega = \frac{i\gamma k^2}{i - \alpha} = \frac{\gamma k^2 (1 - i\alpha)}{1 + \alpha^2}.
\end{equation}
This yields the dispersion relation for the phase slippage magnons,
\begin{equation}
    \omega(k) = \frac{\gamma k^2}{1 + \alpha^2} - i \frac{\gamma \alpha k^2}{1 + \alpha^2}.
\end{equation}

\bibliography{betterbib}

\end{document}